\documentclass[aps,preprint,prl]{revtex4}  
  
\usepackage{amsfonts}  
\usepackage{amsmath}  
\usepackage{amssymb}  
\usepackage{graphicx}  
  
\begin{document}

\title{Sign of refractive index and group velocity in left-handed media}  
\author{A. L. Efros}  
\affiliation{University of Utah, Salt Lake City UT, 84112 USA}  
\author{A. L. Pokrovsky}  
\affiliation{University of Utah, Salt Lake City UT, 84112 USA}  
  
\begin{abstract}
We argue that the widely spread opinion that the left-handed media (LHM)  
are characterized by a negative refractive index $n_-$ is misleading. 
 Since $n$  
does not enter into Maxwell's equations and boundary conditions, any medium   
may be described by both positive $n$ and negative $n_-=-n$. Two thermodynamic 
inequalities are presented, that make a difference
between   
the LHM and the regular media (RM). 
The first one reads that the group  
velocity is positive in the RM and negative in the LHM.   
The second one is that the product ${\rm Re}(n)  {\rm Im}(n)$ is positive 
in the RM and negative in the LHM. 
Both inequalities are invariant with respect to the 
change $n \rightarrow n_-$. 
However, to use $n_-$ one should change some traditional 
electrodynamics definitions.   
\end{abstract}  
  
\maketitle

\section{Introduction}

Victor Veselago was the first to consider the left-handed media (LHM)   
\cite{ve,ve1}, which he defined as the media with simultaneously negative  
and almost real electric permittivity $\epsilon$ and magnetic permeability   
$\mu$ in some frequency range. He showed that the LHM have a number of  
peculiar properties. All these properties follow from the fact that vectors $%
\mathbf{k,E,H}$ of the plane electromagnetic wave form a left-handed  
rather than a right-handed set. 
Then, the Poynting vector $\mathbf{S}$ has a direction  
opposite to the direction of the wave vector $\mathbf{k}$. The left-handed  
set can be easily obtained from Maxwell's equations for harmonic fields $c%
\mathbf{k\times H}=-\omega\epsilon(\omega)\mathbf{E}$ and $c\mathbf{k\times  
E}=\omega \mu(\omega)\mathbf{H}$.  
  
One of the most interesting properties of the LHM is a negative refraction  
at the interface of a regular medium (RM) and a LHM. The negative  
refraction follows immediately from the continuity of   
the tangential component of $\mathbf{k}$ and  
normal component of $\mathbf{S}$.
The anomalous Snell's law which describes negative refraction 
of the wave incident from the RM has a form   
\begin{equation}  
{\frac{\sin\theta_{i}}{\sin\theta_{r}}}=-\frac{n_{l}}{n_{r}},  \label{sn}  
\end{equation}  
where $\theta_{r},\theta_{i}$ are the angles of refraction and incidence  
respectively, $n_{l},n_{r}$ are positive refractive indices of the LHM and  
the RM. The negative refraction has been recently observed by Shelby,  
Smith, and Shultz\cite{sm3} and by Kosaka\cite{kosaka} in completely  
different systems.  
  
Veselago was the first to propose that the refractive index $n_l$ 
in the LHM is negative.  
In our understanding his only reason was that with this
assumption Snell's  
law Eq.(\ref{sn}) acquires the standard form. 
Smith and Kroll (SK) argued\cite{kr}  
that in the LHM the refractive index should be negative while the group 
 velocity (GV) is positive. This point of view is  
widely spread now \cite{nv,va,smsk,p2,mar}. Based upon this statement  
Valanju \textit{et al.}\cite{va} challenged the Veselago's results and   
claimed that the refraction on the RM-LHM interface is positive.   
  
In this paper we critically analyze this point of view. 
First we discuss the definition 
of the refractive index in an isotropic medium. It follows from the wave 
 equation that 
\begin{equation} 
\omega ^{2}n^{2}=c^{2}k^{2}, 
\label{sq} 
\end{equation} 
 where $n^2=\epsilon \mu$. Now one should relate the wave vector {\bf k} 
to $n$.  
The main principle should be that a refractive index represents a 
property of an isotropic medium and it is  
 independent of the direction of vector ${\bf k}$. 
We write the relation in the form 
 \begin{equation} 
{\bf k}=\frac{\omega n}{c} {\bf l}, 
\label{pn} 
\end{equation} 
where ${\bf l}$ is a unit vector in the direction of ${\rm Re}{\bf k}$. By 
 definition ${\rm Re}n=n^{\prime}>0$. Note, that in the isotropic RM the
vector  ${\rm Im}{\bf k}$ has the same direction as ${\rm Re}{\bf k}$.  
Thus, the above definition 
gives that in the RM $n^{\prime}>0$ and $n^{\prime\prime}={\rm Im}(n)>0$.  
The definition of Eq.(\ref{pn}) is  used in Eq.(83.13) of Landau and Lifshitz 
textbook\cite{ll}. We show, however, that in the LHM Eq.(\ref{pn}) 
 leads to a negative $n^{\prime\prime}$.  
 
Another definition, though non-traditional, is as good as the previous one 
for both types of materials. It reads 
 \begin{equation} 
{\bf k}=-\frac{\omega n_-}{c} {\bf l}. 
\label{nn} 
\end{equation} 
In this case $n^{\prime}_-<0$. We argue here that for both types  
of materials all results obtained using definition of Eq.(\ref{pn})  
may be rewritten through $n_-$ by the change $n\rightarrow -n_-$.  
It follows from the fact that neither Maxwell's equation nor the  
boundary conditions include the refractive index.  
 
There is the third definition which goes back to Brillouin\cite{brill}.  
One may consider $n=\sqrt{\mu \epsilon}$ as an analytical  
function of the complex frequency $\omega$ 
in the upper half plane of $\omega$.  
According to Brillouin one 
should choose the Riemann  surface of this function as a  
surface where  $n\rightarrow 1$ at  
$\omega\rightarrow\infty$ along the real axis.  
Brillouin has  shown how to make an analytical continuation to find  
both $n^{\prime}$ and $n^{\prime\prime}$ everywhere along the real axis for  
the case $\mu=1$ and $\epsilon(\omega)$ having an anomalous dispersion.  
In this case $n^{\prime}$ happens to be positive.  
Thus, it does not contradict Eq.(\ref{pn}). 
 
SK show that for some specific form of the product 
$\epsilon(\omega)\mu(\omega)$ the Brillouin's method gives $n^{\prime}<0$ just in the region with both $\epsilon<0,\mu<0$.  
Note, that we are not aware of any general theorem that 
proves this statement for any functions  
 $\epsilon(\omega)$ and $\mu(\omega)$ with usual analytical properties.  
The result of SK means, that one should use the
definition of Eq.(\ref{pn}) for the frequency range where $\epsilon$, $\mu$  
are positive and Eq.(\ref{nn}) where both $\epsilon$, $\mu$ are negative.  
However, from our point of view, both definitions are equivalent. 
SK argued also, that one can obtain the correct sign for the energy losses  
in the LHM only when $n'<0$. We show below that this statement is incorrect.  
 
\section{General theorems}
First we prove two important theorems which are independent of the 
above definitions of the refractive index. 
    

{\it {\bf Theorem 1:} The group velocity $\partial \omega/ \partial {\bf k}$  
in an isotropic medium 
is positive in the RM and negative in the LHM. 
} In fact, this means that vectors $\partial \omega/ \partial {\bf k}$ and ${\bf k}$ 
are parallel in the RM and antiparallel in the LHM.   
  
{\it {\bf Proof:}}  
Taking the gradient of $\omega  
^{2}n^{2}=c^{2}k^{2}$ in $\mathbf{k}$ space one gets  
\begin{equation}  \label{dwdk}  
\frac{\partial \omega }{\partial \mathbf{k}}=\frac{2c^{2}\mathbf{k}}{d\left[  
\omega ^{2}n^{2}\right] /d\omega }.  
\end{equation}%
The total time-averaged electromagnetic energy density of the plane wave is   
\cite{ll}   
\begin{equation}  \label{u}  
\overline{U}=\frac{1}{16\mu \omega }\frac{d\left[ \omega ^{2}n^{2}\right] }{%
d\omega }|E|^{2}.  
\end{equation}  
Since the energy density $\overline{U}$ is positive, one gets that   
\begin{equation}  \label{ineq}  
\frac{d}{d\omega }\left[ \omega ^{2}n^{2}\right] <0 {\rm \ \ \ (for\ the\ LHM)},  
\end{equation}  
\begin{equation}  \label{ineq1}  
\frac{d}{d\omega }\left[ \omega ^{2}n^{2}\right] >0 {\rm \ \ \ (for\ the\ RM)}.  
\end{equation}  
Thus, Eqs.(\ref{dwdk},\ref{ineq},\ref{ineq1}) prove the statement.  
One can see that 
the criterion which distinguishes between LHM and RM is
independent of the sign of $n$.

{\it {\bf Note:}} Eq.(\ref{dwdk}) is valid for both definitions of 
 $n$ given by Eqs(\ref{pn},\ref{nn}). Assuming the definition of Eq.(\ref{pn}),  
one can restore the usual equation for the GV  
\begin{equation}  
\frac{\partial \omega }{\partial \mathbf{k}}=\frac{\mathbf{k}}{k}\,  
\frac{c}{d(n\omega) /d\omega }. 
\label{gvp}  
\end{equation} 
However, this equation is not valid at $n<0$.  
Assuming the definition of Eq.(\ref{nn}), one gets 
\begin{equation}  
\frac{\partial \omega }{\partial \mathbf{k}}=-\frac{\mathbf{k}}{k}\,  
\frac{c}{d(n_- \omega) /d\omega }. 
\label{gvn}  
\end{equation}%
SK claim that GV is positive in the LHM because they use 
Eq.(\ref{gvp}) with non-traditional negative $n$. The most important 
mistake of Valanju \textit{et al.}\cite{va} is of the same origin.

{\it {\bf Corollary:} The main statement of Veselago, that 
the vectors ${\bf S}$ and ${\bf k}$ have opposite directions 
in the LHM, can be obtained from  
the condition that the energy of the electromagnetic field
 is positive without using the   
first order Maxwell's equations.}  
  
{\it {\bf Proof:}}  
It is easy to show that for a  
plane wave in an isotropic nearly lossless medium the time averaged Poynting 
vector is  
\begin{equation}    
\label{s}  
\mathbf{S} = \overline{U} \frac{\partial \omega}{ \partial \mathbf{k}}.
\end{equation}  
  
Thus, using Theorem 1 one can  
obtain the main result of Veselago: in the LHM the vectors \textbf{S}  
and \textbf{k} have opposite directions. The negative RM-LHM refraction  
follows from this result and standard boundary conditions at the  
interface. 
  
{\it {\bf Notes:}} 
1. From Eqs.(\ref{ineq},\ref{ineq1}) follows that the negative sign of  
the GV and the Veselago results, including the 
negative refraction at the RM-LHM interface, are independent of the sign of $n$. 
 
2. In the isotropic medium with small losses  
the negativeness of the GV might be  
a more general signature of the LHM than $\epsilon <0$, $\mu < 0$.

{\it {\bf Theorem 2:} The product $n' n''$ is positive in the  
RM and negative  
in the LHM.}

{\it {\bf Proof:}} This statement based upon the thermodynamic   
inequalities for imaginary parts of $\epsilon$ and $\mu$ in   
a passive material \cite{ll}(p. 274)  
\begin{equation}  
\label{imem}  
\epsilon'' >0, \quad \mu'' >0.  
\end{equation}  
The dispersion equations has a form   
$k^2 = (n'+in'')^2\omega^2/c^2 =   
(\epsilon ' + i \epsilon '')(\mu ' + i \mu '')\, \omega^2/c^2$.   
The imaginary part of this relation reads   
  
\begin{equation}  
\label{em}  
\epsilon' \mu'' + \mu' \epsilon'' = 2 n' n''.  
\end{equation}  
Therefore    
\begin{equation}  
\label{n2p}  
n' n'' >0 {\rm\ \  in\ the\ RM,} \quad n' n'' <0 {\rm\ \  in\ the\ LHM.}   
\end{equation}

\section{Sign of refractive index}
Thus, using  the fundamental thermodynamic properties we have shown, that  
the sign of the product $n'n''$ and the sign of the GV  
are different for the LHM and RM. Both  $n'n''$ and the GV are invariant  
with respect to change of the sign of $n$ and  
in the rest of our paper we  
discuss whether or not the sign of $n$ has any physical meaning. 
  
\begin{figure}  
\includegraphics[width=8.6cm]{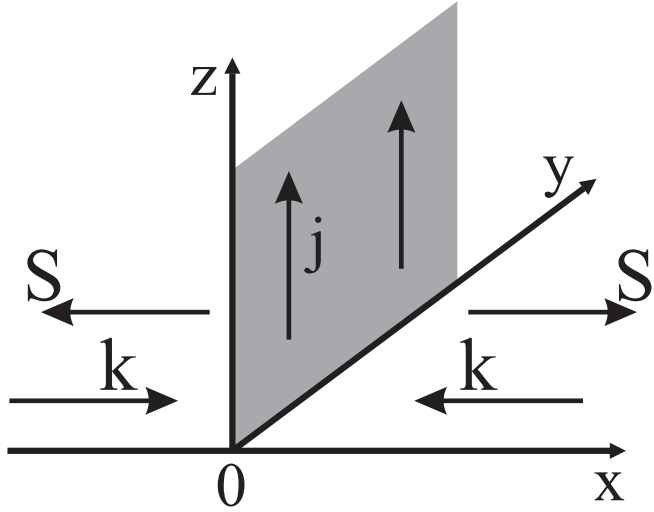} \vspace{0.5cm}  
\caption{An infinite sheet of current in the $y-z$ plane, embedded into   
  the LHM. ${\bf S}$, ${\bf k}$ and ${\bf j}$ show the directions   
  of the Poynting vector,  
  the wave vector and the current respectively.}  
\label{fig1}  
\end{figure}

To illustrate our point of view we consider the same problem as SK.  
They study an infinite sheet of current at $x=0$, embedded into the LHM   
(Fig. \ref{fig1}). The surface current density is $j=j_{0}\exp(-i\omega t)$.  
They solve the equation for the $z$-component of the electric field $E(x,t)$   
\begin{equation}  
\frac{\partial^{2}E(x)}{\partial x^{2}}+\frac{\omega^{2}}{c^{2}}n^{2}E(x)=-%
\frac{4\pi i\omega}{c^{2}}\mu j_{0}\delta(x)  
\end{equation}  
to show that the power radiated by the sheet of current is positive only if  
the refractive index is negative.

Now we demonstrate that the same condition can be fulfilled without   
introducing the negative $n$.   
Following Veselago we choose the direction of the wave vector  
${\bf k}$ from the causality condition. In the LHM it should be directed towards the  
source of the wave, while in the RM it should be directed outward the source.  
In both, LHM and RM ${\bf S}$ should be directed outward the source   
(Fig.\ref{fig1}).
Using boundary condition at $x=0$   
\begin{equation}  
\left.\frac{\partial E(x)}{\partial x}\right|_{+}-\left.\frac{\partial E(x)}{%
\partial x}\right|_{-}=-\frac{4\pi i\omega}{c^{2}}\mu j_{0},  
\end{equation}  
the electric field can be written in the form   
\begin{equation}  
E(x,t)=\frac{2\pi\omega}{k c^{2}}\mu j_{0}e^{i({\bf k}\cdot {\bf x}-\omega t)} = 
\frac{2\pi\omega}{k c^{2}}\mu j_{0}e^{i(- k|x|-\omega t)},  
\end{equation}  
where ${\bf k} = k {\bf l}$, ${\bf l}$ is a unit vector in the direction of  
${\rm Re}{\bf k}$ (see Fig.\ref{fig1}). Taking into account that $\omega n = c k$, and introducing real  
and imaginary parts of the refractive index the solution can be written in the  
following form 
\begin{equation}  
\label{001} 
E(x,t)=\frac{2\pi\omega}{n \kappa c^{2}}\mu j_{0}e^{- (i n' - n'')\kappa |x|- i \omega t},  
\end{equation}
where $\kappa=\omega/c>0$. From Eq.(\ref{001}) one can see 
that  $n''$ should be negative in the LHM to 
provide the decay of the wave in the direction of the energy propagation. 
Since we have chosen $n'>0$ this result follows also from Theorem 2.

Using the solution for the electric field one can see that   
\begin{equation}  
-\frac{1}{2}\int j^{\ast}E(x)dx=-\frac{\pi\omega\mu}{k c^{2}}j_{0}^{2}=%
\frac{\pi|\mu|}{cn}j_{0}^{2}>0.  
\end{equation}  
The refractive index $n$ obeys the usual equation $n=c k/\omega $. It  
is easy to show that $S_{x}=\pi |\mu |j_{0}^{2}/2cn>0$ at $x>0$ and that   
$S_{x}=-\pi |\mu |j_{0}^{2}/2cn<0$ at $x<0$.  
  
Now we analyze solution of SK. It reads   
\begin{equation}    
\label{esmkr}  
E(x,t)=-\frac{2\pi\omega}{n_{-}\kappa c^{2}}\mu j_{0}e^{i(n_{-}\kappa  
|x|-\omega t)}. 
\end{equation}     
With $n_- = n_-' + in_-''$ this equation takes the form 
\begin{equation}    
\label{esmkr1}  
E(x,t)=-\frac{2\pi\omega}{n_{-}\kappa c^{2}}\mu j_{0}e^{(i n_{-}'-n_-'')\kappa  
|x|-i\omega t}. 
\end{equation} 
It follows then, that the solution written in this form dictates
$n_-''$ to be positive.  
This result also follows from Theorem 2 for the LHM because $n'_-$ is negative.

Using expression Eq.(\ref{esmkr}) one gets   
\begin{equation}  
-\frac{1}{2}\int j^{\ast}E(x)dx=\frac{\pi\mu}{cn_{-}}j_{0}^{2} =  
\frac{\pi|\mu|}{cn}j_{0}^{2}>0.  
\end{equation}  
Thus, we see that the descriptions of this problem for the LHM  
in terms of $n$ and $n_{-}$ are identical. 
The SK's solution Eq.(\ref{esmkr1}) can be obtained from
Eq.(\ref{001}) by changing
$n\rightarrow -n_-$.
 
Moreover, even the RM can be  
described in terms of negative refractive index  $n_{-}$.   
Suppose, that our sheet of current is embedded into a RM.   
Then the correct solution for  
the electric field can be written in the form   
\begin{equation}    
\label{erhm}  
E(x,t)=\frac{2\pi\omega}{n_{-}\kappa c^{2}}\mu j_{0}e^{i(-n_{-}\kappa  
|x|-\omega t)}.  
\end{equation}  
Note that this time $n_-''$ should be negative, that also supports our theorem 
because we consider the RM with negative $n_-'$. 
 
The above examples confirm our point of view that the sign of $n'$ is 
 not the criterion which makes a difference between the  LHM and RM. We 
think that these examples reflect the  general case and the reason is that any electrodynamics 
 problem can be formulated in terms of $\epsilon$ and $\mu$, so that the question  
of the sign of $n$ does not appear.

Since the mathematical description  
with both $n$ and $n_{-}$ is the same, we doubt that there is a way to  
find the sign of $n$ either from experimental or from computational data.

Smith \textit{et al.}\cite{smsk} attempted to  
find the sign of $n$ from the reflection and transmission. 
Transmission $t$ and reflection $r$ coefficients for waves normally incident 
on a LHM slab with the width $d$ can be written in a form  
\begin{equation}  
t^{-1}=\left[\cos{(n \kappa d)} - \frac{i}{2}\left( \mu + \epsilon\right)  
\frac{\sin{(n\kappa d)}}{n} \right]e^{i\kappa d}  
\end{equation}  
\begin{equation}  
r = -\frac{i}{2} t e^{i\kappa d} (\mu - \epsilon)\frac{\sin{(n\kappa d)}}{n}.  
\end{equation}  
 
These equations are invariant with respect to change $n\rightarrow -n_-$. 
However the authors of \cite{smsk} solve them with respect to $n'$,$n''$ and impose  
the condition $n''>0$. Then, in agreement with Theorem 2, 
they find that in the LHM $n'<0$. 
They claim that the condition $n''>0$ is valid for any passive 
 material. From our point of view the conditions 
${\rm Im}\epsilon>0$ and ${\rm Im}\mu>0$
define a passive material, and we are not aware of 
any theorem about the sign of $n''$.

\section{Fermat's principle for the LHM} 
\begin{figure}  
\includegraphics[width=8.6cm]{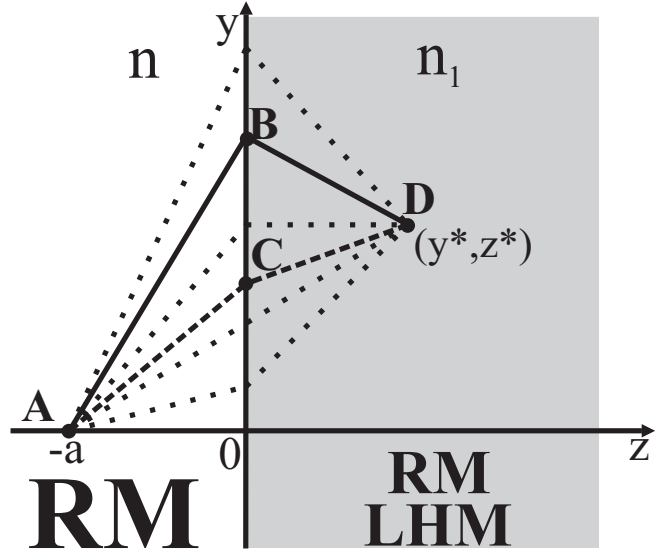} \vspace{0.5cm}  
\caption{Different paths of rays traveling from point A to point D. 
If the medium at $z>0$ is a LHM, the path ABD obeys the Fermat's principle. 
It shows negative refraction. 
In the case when both media are RM the path ACD, which has a positive 
refraction, obeys the Fermat's principle.}  
\label{fig2}  
\end{figure}

Fermat's principle for the LHM has been recently formulated
by Veselago \cite{vefe} in terms of negative refractive index.
Here we show how to do it keeping the refractive index positive for both 
LHM and RM.
A simple generalization of the Maupertuis principle 
for the particles reads
\begin{equation}
\label{fermat} 
\delta \int_{{\bf r}_1}^{{\bf r}_2} \mathbf{k}\cdot d\mathbf{l} = 0, 
\end{equation}
where $d{\bf l}$ is a vector in the direction of propagation
of photons which is parallel to the GV or to the Poynting vector.
In the RM ${\bf k}\cdot d{\bf l} = (\omega n/c) dl$, while in the LHM
${\bf k}\cdot d{\bf l} = -(\omega n/c)dl$.
Equation(\ref{fermat}) describes the propagation of rays.
Following Veselago\cite{vefe} we apply this principle to refraction
of the waves at the RM-RM and RM-LHM interfaces. 
Consider a point A with coordinates ($0,-a$) in a RM with the
refractive index $n$ and a point D with coordinates ($y^*, z^*$) in
the material with the positive refractive index $n_1$ (Fig. \ref{fig2}). 
We consider two cases:
first - when the media at $z>0$ is a RM, second - when it is a LHM described 
by a positive refractive index $n_1$.
Using Fermat's principle we find the path of the 
ray of light coming from A to D.
Equation (\ref{fermat}) in the case of RM-RM interface has a form 
\begin{equation}
\label{frm}
\frac{ny}{\sqrt{a^2+y^2}}+\frac{n_1 (y-y^*)}{\sqrt{(z^*)^2+(y-y^*)^2}}=0.
\end{equation} 
This equation gives the usual Snell's law
$n \sin{\theta_i} = n_1 \sin{\theta_r}$ because
$\sin{\theta_i} = y/\sqrt{a^2+y^2}$, 
$\sin{\theta_r} = (y^*-y)/\sqrt{(z^*)^2+(y^*-y)^2}$.
The propagation of rays in this case is shown at Fig. \ref{fig2} 
by the dashed line ACD.

In the case of RM-LHM interface the Fermat's principle Eq.(\ref{fermat}) gives 
\begin{equation}
\label{flhm}
\frac{ny}{\sqrt{a^2+y^2}}-\frac{n_1 (y-y^*)}{\sqrt{(z^*)^2+(y-y^*)^2}}=0,
\end{equation} 
which is the anomalous Snell's law $n \sin{\theta_i} = -n_1 \sin{\theta_r}$
providing the negative refraction. 
The propagation of rays 
is shown by the solid line ABD at Fig. \ref{fig2}.
Note, that $n=n_1$ is a special case.
Under this condition there is a focal point inside the LHM with coordinates
$(0, a)$. 
One can see from Eq.(\ref{flhm}) that at $n=n_1$ all rays 
go through this point.
The optical length of the path, which can be defined as 
\begin{equation}
\frac{c}{\omega} \int_{A}^{D} \mathbf{k}\cdot d\mathbf{l},
\end{equation}
is zero for any ray going from A to the focal point.
The focal point is absent for $n \neq n_1$ \cite{ve}.

\section{Conclusions}

1. We have proven two theorems based upon fundamental 
thermodynamic inequalities  
that show the difference between the LHM and the RM. 
The first one states that the GV is negative in the LHM and positive in the  
RM. If the losses are small, one could get all Veselago's results  
from this theorem without using the first order Maxwell's equations.  
The second theorem claims that the product $n'n''$ is positive in the RM and 
negative in the LHM. 
 
2. Both theorems are invariant with respect to change of 
the sign of refractive  
index.  
 
3. We have shown that the introduction of negative refractive index is  
possible in both RM and LHM. However it may be misleading because some  
traditional electrodynamics expressions including Eq.(\ref{gvp}) 
become incorrect 
if $n$ is negative. That is the reason why some authors\cite{kr,va} claim that 
GV in the LHM is positive.  

4. We have generalized the Fermat's principle for the LHM 
using positive value of $n$ and have shown that it leads to a negative
refraction at the RM-LHM interface.  

5. Thus we do not see any reason to ascribe a negative refractive index  
to the LHM.

\begin{acknowledgments}  
The work has been funded by the NSF grant DMR-0102964.  
\end{acknowledgments}  
  
\bibliographystyle{apsrev}  
\bibliography{index}

\end{document}